\documentclass[conference]{IEEEtran}
\usepackage{graphicx}
\usepackage{amsthm}
\usepackage{epsfig}
\usepackage{latexsym}
\usepackage{amsfonts}
\usepackage{here}
\usepackage{rawfonts}
\usepackage[latin1]{inputenc}
\usepackage[T1]{fontenc}
\usepackage{calc}
\usepackage{capitalgreekitalic}
\usepackage{url}
\usepackage{enumerate}
\usepackage{color}
\usepackage[tbtags]{amsmath}
\usepackage{amssymb}
\usepackage{upref}
\usepackage{epic,eepic}
\usepackage{times}
\usepackage{dsfont}
\usepackage{comment}
\usepackage{cite}
\usepackage{bbm} 
\usepackage{multirow}

\usepackage{dsfont}

\newcounter{step}
\newlength{\totlinewidth}
  {\end{list}%
  \rule{\linewidth}{1pt}}
\newcounter{substep}

  {\end{list}}

\newlength{\aligntop}
\newlength{\alignbot}
 

\IEEEoverridecommandlockouts

\usepackage{algorithm}
\usepackage{algorithmic}

\usepackage{subfigure}

\begin{document}
\title{\huge A Vision of 6G Wireless Systems: Applications, Trends, Technologies, and Open Research Problems\vspace*{-0.4em}}
\author{{Walid Saad\IEEEauthorrefmark{1}}, Mehdi Bennis\IEEEauthorrefmark{2}, and Mingzhe Chen\IEEEauthorrefmark{3}$^,$\IEEEauthorrefmark{4}$^,$\IEEEauthorrefmark{1} 
 \vspace*{0em}\\ 
\IEEEauthorrefmark{1}\small Wireless@VT, Bradley Department of Electrical and Computer Engineering, Virginia Tech, Blacksburg, VA, USA, Email: \protect\url{walids@vt.edu}.\\
\IEEEauthorrefmark{2}\small CWC - Centre for Wireless Communications, University of Oulu, Finland, Email: \protect\url{mehdi.bennis@oulu.fi}.\\
\authorblockA{\small \IEEEauthorrefmark{3}The Future Network of Intelligence Institute, The Chinese University of Hong Kong, Shenzhen, China, Email: \protect\url{chenmingzhe@cuhk.edu.cn}. \\
\IEEEauthorrefmark{4}Department of Electrical Engineering, Princeton University, Princeton, NJ, USA.\\
\thanks{{This research was supported by the U.S. National Science Foundation under Grant CNS-1836802.}}
}\vspace*{-3em}
 }

\maketitle

\vspace{0cm}
\begin{abstract}
The ongoing deployment of 5G cellular systems is continuously exposing the inherent limitations of this system, compared to its original premise as an enabler for Internet of Everything applications. These 5G drawbacks are spurring worldwide activities focused on defining the next-generation 6G wireless system that can truly integrate far-reaching applications ranging from autonomous systems to extended reality. Despite recent 6G initiatives\footnote{One example is the 6Genesis project in Finland (see https://www.oulu.fi/6gflagship/).}, the fundamental architectural and performance components of 6G remain largely undefined. In this paper, we present a holistic, forward-looking vision that defines the tenets of a 6G system. We opine that 6G will not be a mere exploration of more spectrum at high-frequency bands, but it will rather be a convergence of upcoming technological trends driven by exciting, underlying services. In this regard, we first identify the primary drivers of 6G systems, in terms of applications and accompanying technological trends. Then, we propose a new set of service classes and expose their target 6G performance requirements. We then identify the enabling technologies for the introduced 6G services and outline a comprehensive research agenda that leverages those technologies. We conclude by providing concrete recommendations for the roadmap toward 6G. Ultimately, the intent of this article is to serve as a basis for stimulating more out-of-the-box research around 6G.
\end{abstract}
%
%
%
%
%
\section{Introduction} \label{se:1}
To date, the wireless network evolution was primarily driven by a need for higher rates, which mandated a continuous 1000x increase in network capacity. While this demand for wireless capacity will continue to grow, the emergence of the Internet of Everything (IoE) system, connecting millions of people and billions of machines, is yielding a radical paradigm shift from the rate-centric enhanced mobile broadband (eMBB) services of yesteryears towards ultra-reliable, low latency communications (URLLC).

Although the fifth generation (5G) cellular system was marketed as the key IoE enabler, through concerted 5G standardization efforts that led to the first 5G new radio (5G NR) milestone and subsequent 3GPP releases, the initial premise of 5G -- as a true carrier of IoE services -- is yet to be realized. One can argue that the \emph{evolutionary} part of 5G (i.e., supporting rate-hungry eMBB services) has gained significant momentum, however, the promised \emph{revolutionary} outlook of 5G -- a system operating almost exclusively at high-frequency millimeter wave (mmWave) frequencies and enabling  heterogeneous IoE services -- has thus far remained a mirage. Although the 5G systems that are currently being marketed will readily support basic IoE and URLLC services (e.g., factory automation), it is debatable whether they can deliver tomorrow's smart city IoE applications. Moreover, although 5G will eventually support fixed-access at mmWave frequencies, it is more likely that early 5G roll-outs will still use sub-6 GHz for supporting mobility. 

Meanwhile, an unprecedented proliferation of new IoE services is ongoing. Examples range from extended reality (XR) services (encompassing augmented, mixed, and virtual reality (AR/MR/VR)) to telemedicine, haptics, flying vehicles, brain-computer interfaces, and connected autonomous systems. These applications will disrupt the original 5G goal of supporting short-packet, sensing-based URLLC services. To successfully operate IoE services such as XR and connected autonomous systems, a wireless system must simultaneously deliver high reliability, low latency, and high data rates, for heterogeneous devices, across uplink and downlink. Emerging IoE services will also require an \emph{end-to-end co-design of communication, control, and computing} functionalities, which to date has been largely overlooked. To cater for this new breed of services, unique challenges must be addressed ranging from characterizing the fundamental rate-reliability-latency tradeoffs governing their performance to exploiting frequencies beyond sub-6 GHz and transforming wireless systems into a self-sustaining, intelligent network fabric which flexibly provisions and orchestrates communication-computing-control-localization-sensing resources tailored to the requisite IoE scenario.

To overcome these challenges, a disruptive \emph{sixth generation (6G)} wireless system, whose design is inherently tailored to the performance requirements of IoE applications and their accompanying technological trends, is needed. The drivers of 6G will be a confluence of past trends (e.g., densification, higher rates, and massive antennas) and of emerging trends that include new services and the recent revolution in wireless devices (e.g., smart wearables, implants, XR devices, etc.), artificial intelligence (AI) \cite{Chen-AI}, computing, and sensing.

\begin{figure*}[!t]
  \begin{center}
   \vspace{0cm}
    \includegraphics[width=15cm]{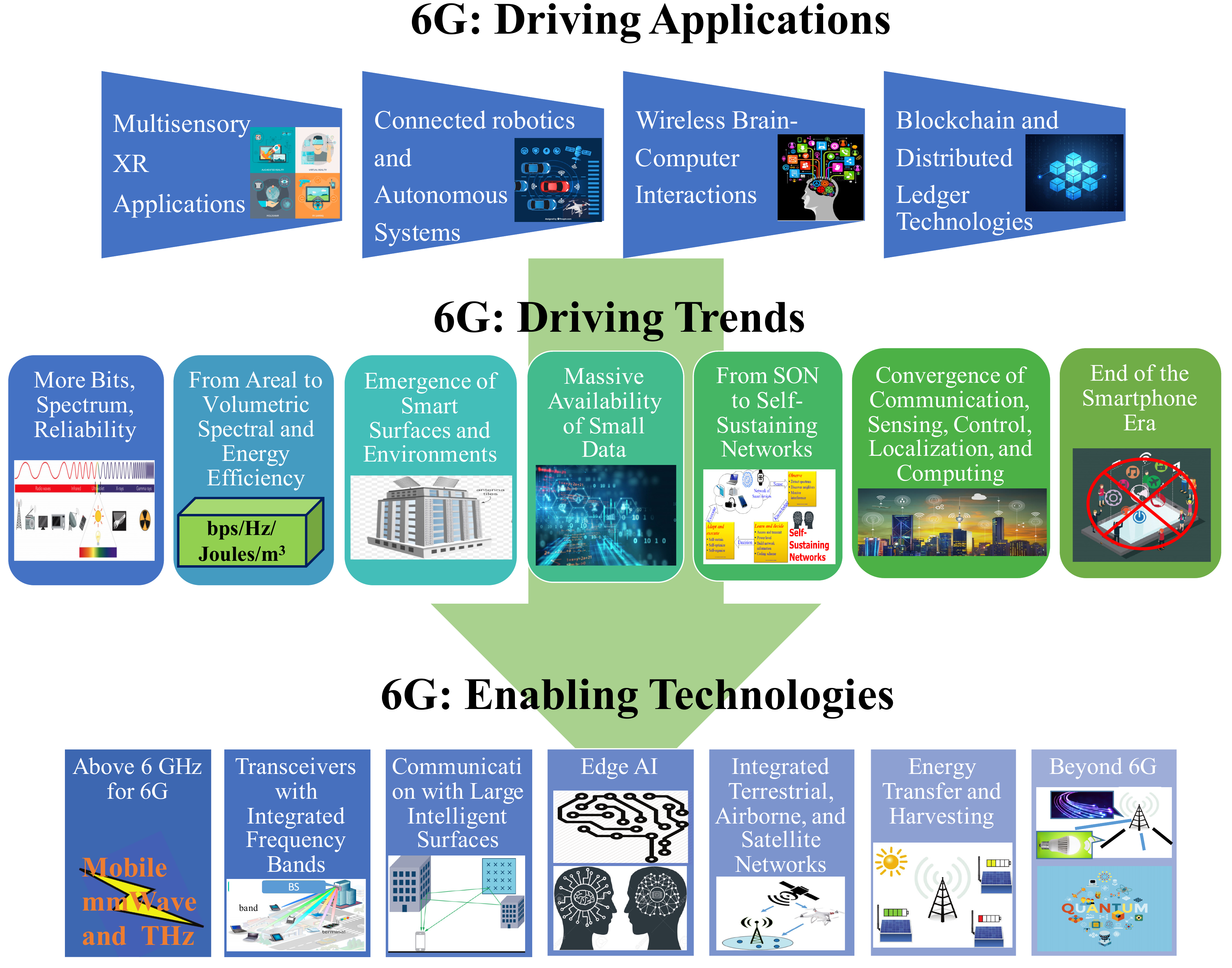}
    \vspace{-0.2cm}
    \caption{\label{figure1} 6G vision: Applications, trends, and technologies.}
  \end{center}\vspace{-0.7cm}
\end{figure*}

The main contribution of this article is a bold, forward-looking vision of 6G systems (see Fig. \ref{figure1}) that identifies the applications, trends, performance metrics, and disruptive technologies, that will drive the 6G revolution. This vision will then delineate new 6G services and provide a concrete research roadmap and recommendations to facilitate the leap from current 5G systems towards 6G. 


\section{6G Driving Applications, Metrics, and New Service Classes}\label{se:2}
Every new cellular generation is driven by innovative applications. 6G is no exception: It will be borne out of an unparalleled emergence of exciting new applications and technological trends that will shape its performance targets while radically redefining standard 5G services. Next, we first introduce the main applications that motivate 6G deployment and, then, discuss ensuing technological trends, target performance metrics, and new service requirements.

\subsection{\color{black}Driving Applications behind 6G and Their Requirements}\label{se:21}
While traditional applications, such as live multimedia streaming, will remain central to 6G, the key determinants of the system performance will be four new application domains:
\subsubsection{\bf Multisensory XR Applications} XR will yield many killer applications for 6G across the AR/MR/VR spectrum. Upcoming 5G systems still fall short of providing a full immersive XR experience capturing all sensory inputs due to their inability to deliver very low latencies for data-rate intensive XR applications. A truly immersive AR/MR/VR experience requires a joint design integrating not only engineering (wireless, computing, storage) requirements but also \emph{perceptual} requirements stemming from human senses, cognition, and physiology.  Minimal and maximal perceptual requirements and limits must be factored into the engineering process (computing, processing, etc.). To do so, a new concept of {\emph{quality-of-physical-experience (QoPE)}} measure is needed to merge physical factors from the human user itself with classical QoS (e.g., latency and rate) and QoE (e.g., mean-opinion score) inputs. Some factors that affect QoPE include brain cognition, body physiology, and gestures. As an example, in \cite{kasgari2018human}, we have shown that the human brain may not be able to distinguish between different latency measures, within the URLLC regime. Meanwhile, in \cite{MehdiEdge}, we showed that visual and haptic perceptions are key for maximizing resource utilization. Concisely, the requirements of XR services are a blend of traditional URLLC and eMBB with incorporated perceptual factors that 6G must support.

\subsubsection{\bf Connected Robotics and Autonomous Systems (CRAS)} A primary driver behind 6G systems is the imminent deployment of CRAS including drone-delivery systems, autonomous cars, autonomous drone swarms, vehicle platoons, and autonomous robotics. The introduction of CRAS over the cellular domain is not a simple case of ``yet another short packet uplink IoE service''. Instead, CRAS mandate control system-driven latency requirements as well as the potential need for eMBB transmissions of high definition (HD) maps. The \emph{notion of QoPE applies once again for CRAS; however, the physical environment is now a control system}, potentially augmented with AI. CRAS are perhaps a prime use case that requires stringent requirements across the  rate-reliability-latency spectrum; a balance that is not yet available in 5G.

\subsubsection{\bf Wireless Brain-Computer Interactions (BCI)} Beyond XR, tailoring wireless systems to their human user is mandatory to support services with direct BCI.  Traditionally, BCI applications were limited to healthcare scenarios in which humans can control prosthetic limbs or neighboring computing devices using brain implants. However, the recent advent of wireless brain-computer interfaces and implants will revolutionize this field and introduce new use-case scenarios that require 6G connectivity. Such scenarios range from enabling brain-controlled movie input to fully-fledged multi-brain-controlled cinema \cite{Brain-Movies-2018}. Using wireless BCI technologies, instead of smartphones, people will interact with their environment and other people using discrete devices, some worn, some implanted, and some embedded in the world around them. This will allow individuals to control their environments through gestures and communicate with loved ones through haptic messages. Such \emph{empathic and haptic communications}, coupled with related ideas such as affective computing in which emotion-driven devices can match their functions to their user's mood, constitute important 6G use cases. Wireless BCI services require fundamentally different performance metrics compared to what 5G delivers. Similar to XR, wireless BCI services need high rates, ultra low latency, and high reliability. However, they are much more sensitive than XR to physical perceptions and necessitate QoPE guarantees.

\subsubsection{\bf Blockchain and Distributed Ledger Technologies (DLT)} Blockchains and DLT will be one of the most disruptive IoE technologies.  Blockchain and DLT applications can be viewed as the next-generation of distributed sensing services whose need for connectivity will require a synergistic mix of URLLC and massive machine type communications (mMTC) to guarantee low-latency, reliable connectivity, and scalability.

\subsection{6G: Driving Trends and Performance Metrics}\label{se:22}
The applications of Section \ref{se:21} lead to new system-wide trends that will set the goals for 6G:

\begin{itemize}
\item {\bf Trend 1 -- More Bits, More spectrum, More Reliability:} Most of the driving applications of 6G require higher bit rates than 5G. To cater for applications such as XR and BCI, 6G must deliver yet another 1000x increase in data rates yielding a target of around 1 Terabit/second. This motivates a need for more spectrum resources, hence prompting further exploration of frequencies beyond sub-6 GHz. Meanwhile, the need for higher reliability will be pervasive across most 6G applications and will be more challenging to meet at high frequencies.

\item {\bf Trend 2 -- From Areal to Volumetric Spectral and Energy Efficiency:} 6G must deal with ground and aerial users, encompassing smartphones and XR/BCI devices along with flying vehicles. This 3D nature of 6G requires an evolution towards a volumetric rather than spatial {\color{black}{({areal})}} bandwidth definition. We envision that 6G systems must deliver high spectral and energy efficiency (SEE) requirements  measured in bps/Hz/\textrm{m$^3$}/Joules. This is a natural evolution that started from 2G (bps) to 3G (bps/Hz), then 4G (bps/Hz/{m$^2$}) to 5G (bps/Hz/m$^2$/Joules).

\item {\bf Trend 3 -- Emergence of Smart Surfaces and Environments:} Current and past cellular systems used base stations (of different sizes and forms) for transmission. We are witnessing a revolution in electromagnetically active surfaces (e.g., using metamaterials) that include man-made structures such as walls, roads, and even entire buildings, as exemplified by the Berkeley ewallpaper project\footnote{See https://bwrc.eecs.berkeley.edu/projects/5605/ewallpaper.}. The use of such smart large intelligent surfaces and environments for wireless communications will drive the 6G architectural evolution.

\item {\bf Trend 4 -- Massive Availability of Small Data:} The data revolution will continue in the near future and shift from centralized, big data, towards massive, distributed ``small'' data. 6G systems must harness both big and small datasets across their infrastructure to enhance network functions and provide new services. This trend motivates new machine learning techniques that go beyond classical big data analytics.

\item {\bf Trend 5 -- From Self-Organizing Networks (SON) to Self-Sustaining Networks:} SON has only been scarcely integrated into 4G/5G networks due to a lack of real-world need. However, CRAS and DLT technologies motivate an immediate need for intelligent SON to manage network operations, resources, and optimization. 6G will require a paradigm shift from classical SON, whereby the network merely adapts its functions to specific environment states, into a \emph{self-sustaining network (SSN)} that can maintain its key performance indicators (KPIs), \emph{in perpetuity}, under highly dynamic and complex environments stemming from the rich 6G application domains. SSNs must be able to not only adapt their functions but to also sustain their resource usage and management (e.g., by harvesting energy and exploiting spectrum) to autonomously maintain high, long-term KPIs. SSN functions must leverage the recent revolution in AI technologies to create AI-powered 6G SSNs.

\item {\bf Trend 6 -- Convergence of Communications, Computing, Control, Localization, and Sensing (3CLS):} The past five generations of cellular systems had one exclusive function: wireless communications. {\color{black}However, 6G will disrupt this premise through a convergence (i.e., joint and simultaneous offering) of various functions that include communications, computing \cite{8265188}, control, localization, and sensing.} We envision 6G as a multi-purpose system that can deliver multiple 3CLS services which are particularly appealing and even necessary for applications such as XR, CRAS, and DLT where tracking, control, localization, and computing are an inherent feature. Moreover, sensing services will enable 6G systems to provide users with a \emph{3D mapping of the radio environment} across different frequencies. Hence, 6G systems must tightly integrate and manage 3CLS functions. {\color{black}Note that, the evolutions pertaining to previous trends will gradually enable 6G systems to readily provide 3CLS.}

\item {\bf Trend 7 -- End of the Smartphone Era:} Smartphones were central to 4G and 5G. However, recent years witnessed an increase in wearable devices whose functionalities are gradually replacing those of smartphones. This trend is further fueled by applications such as XR and BCI. The devices associated with those applications range from smart wearables to integrated headsets and smart body implants that can take direct sensory inputs from human senses; bringing an end to smartphones and potentially driving a majority of 6G use cases.

\end{itemize}
As shown in Table \ref{ta:1}, collectively, these trends impose new performance targets and requirements that will be met in two stages: a) A beyond 5G evolution and b) A revolutionary 6G step.
 \begin{table*}
\centering
  \newcommand{\tabincell}[2]{\begin{tabular}{@{}#1@{}}#2.4\end{tabular}}
\renewcommand\arraystretch{1.5}
 \caption{
    \vspace*{-0.05em}Requirements of 5G vs. Beyond 5G vs. 6G.}\label{ta:1}\vspace*{-0.5em}
\centering
\begin{tabular}{|c|l|l|l|}
\hline
 \multicolumn{1}{|c|}{\multirow{2}{*}{\textbf{}}}&\multicolumn{1}{|c|}{\multirow{2}{*}{\textbf{5G}}}   &  \multicolumn{1}{|c|}{\multirow{2}{*}{\textbf{ Beyond 5G}}} &  \multicolumn{1}{|c|}{\multirow{2}{*}{\textbf{ 6G}}} \\ 
 &&&\\
\hline
\multirow{5}{*}{{\bf Application Types }}& \multirow{1}{*}{{$\bullet$ eMBB.}} &\multirow{1}{*}{$\bullet$ Reliable eMBB.} &\multirow{1}{*}{New applications (see Section \ref{se:23}):} \\
&$\bullet$ URLLC.&$\bullet$ URLLC. &$\bullet$ MBRLLC.\\
&$\bullet$ mMTC.&$\bullet$ mMTC. & \multirow{1}{*}{$\bullet$ mURLLC.}\\ 
&&$\bullet$ Hybrid (URLLC + eMBB).&$\bullet$ HCS.\\
&&&$\bullet$ MPS. \\
\hline

\multirow{4}{*}{{\bf Device Types}}& \multirow{1}{*}{{$\bullet$ Smartphones.}} &\multirow{1}{*}{$\bullet$ Smartphones.} &\multirow{1}{*}{$\bullet$ Sensors and DLT devices.} \\
&$\bullet$ Sensors.&$\bullet$ Sensors.&$\bullet$ CRAS.\\
&$\bullet$ Drones.&$\bullet$ Drones. &$\bullet$ XR and BCI equipment. \\ 
&&\multirow{1}{*}{$\bullet$ XR equipment.} &$\bullet$ Smart implants. \\ 

\hline
\multirow{4}{3cm}{{\bf Spectral and Energy Efficiency Gains\footnotemark[3] with Respect to Today's Networks}}& \multirow{4}{*}{{10x in bps/Hz/m$^2$/Joules}}  & \multirow{4}{*}{{100x in bps/Hz/m$^2$/Joules}}& \multirow{4}{*}{{1000x in bps/Hz/m$^3$/Joules (volumetric)}}\\
&&&\\
&&&\\
&&&\\
\hline
\multirow{2}{3cm}{{\bf Rate Requirements}}& \multirow{2}{*}{{ 1 Gbps}}  & \multirow{2}{*}{{100 Gbps}}& \multirow{2}{*}{{1 Tbps}}\\
&&&\\
\hline
\multirow{2}{3cm}{{\bf End-to-End Delay Requirements}}& \multirow{2}{*}{{ 5 ms}}  & \multirow{2}{*}{{1 ms}}& \multirow{2}{*}{{< 1 ms}}\\
&&&\\
\hline
\multirow{2}{3cm}{{\bf Radio-Only Delay Requirements}}& \multirow{2}{*}{{ 100 ns}}  & \multirow{2}{*}{{100 ns}}& \multirow{2}{*}{{10 ns}}\\
&&&\\
\hline
\multirow{2}{3cm}{{\bf Processing Delay}}& \multirow{2}{*}{{ 100 ns}}  & \multirow{2}{*}{{50 ns}}& \multirow{2}{*}{{10 ns}}\\
&&&\\
\hline
\multirow{2}{3cm}{{\bf End-to-End Reliability Requirements}}& \multirow{2}{*}{{ \color{black}99.999\%}}  & \multirow{2}{*}{{\color{black}99.9999\%}}& \multirow{2}{*}{{\color{black}99.99999\%}}\\
&&&\\
\hline
\multirow{3}{*}{{\bf Frequency Bands}}& \multirow{1}{*}{{$\bullet$ Sub-6 GHz.}} &\multirow{1}{*}{$\bullet$ Sub-6 GHz.} &\multirow{1}{*}{$\bullet$ Sub-6 GHz.} \\
&$\bullet$ MmWave for fixed access.&\multirow{2}{5cm}{$\bullet$ MmWave for fixed access at 26 GHz and 28GHz.}&$\bullet$ MmWave for mobile access.\\
&& &$\bullet$ Exploration of THz bands (above 300 GHz). \\ 
&& &$\bullet$ Non-RF (e.g., optical, VLC, etc.). \\ 
\hline
\multirow{6}{*}{{\bf Architecture}}& \multirow{3}{3cm}{{$\bullet$ Dense sub-6 GHz small base stations with umbrella macro base stations.}} &\multirow{2}{5cm}{$\bullet$ Denser sub-6 GHz small cells with umbrella macro base stations.} &\multirow{3}{4cm}{$\bullet$ Cell-free smart surfaces at high frequency supported by mmWave tiny cells for mobile and fixed access.} \\
&&&  \\
&&$\bullet$ < 100 m tiny and dense mmWave cells.&\\ 
&\multirow{3}{3cm}{{$\bullet$  MmWave small cells of about 100 m (for fixed access).}} & &\multirow{2}{4cm}{$\bullet$ Temporary hotspots served by drone-carried base stations or tethered balloons.}\\
&&&\\
&&&$\bullet$ Trials of tiny THz cells.\\
\hline
\end{tabular}
 \vspace{-0.2cm}

\end{table*}

\subsection{New 6G Service Classes}\label{se:23}

Beyond imposing new performance metrics, the new technological trends will redefine 5G application types by morphing classical URLLC, eMBB, and mMTC and introducing new services (summarized in Table \ref{ta:2}):
\subsubsection{\bf{Mobile Broadband Reliable Low Latency Communication}} 
\begin{figure*}[!t]
  \begin{center}
   \vspace{0cm}
    \includegraphics[width=15cm]{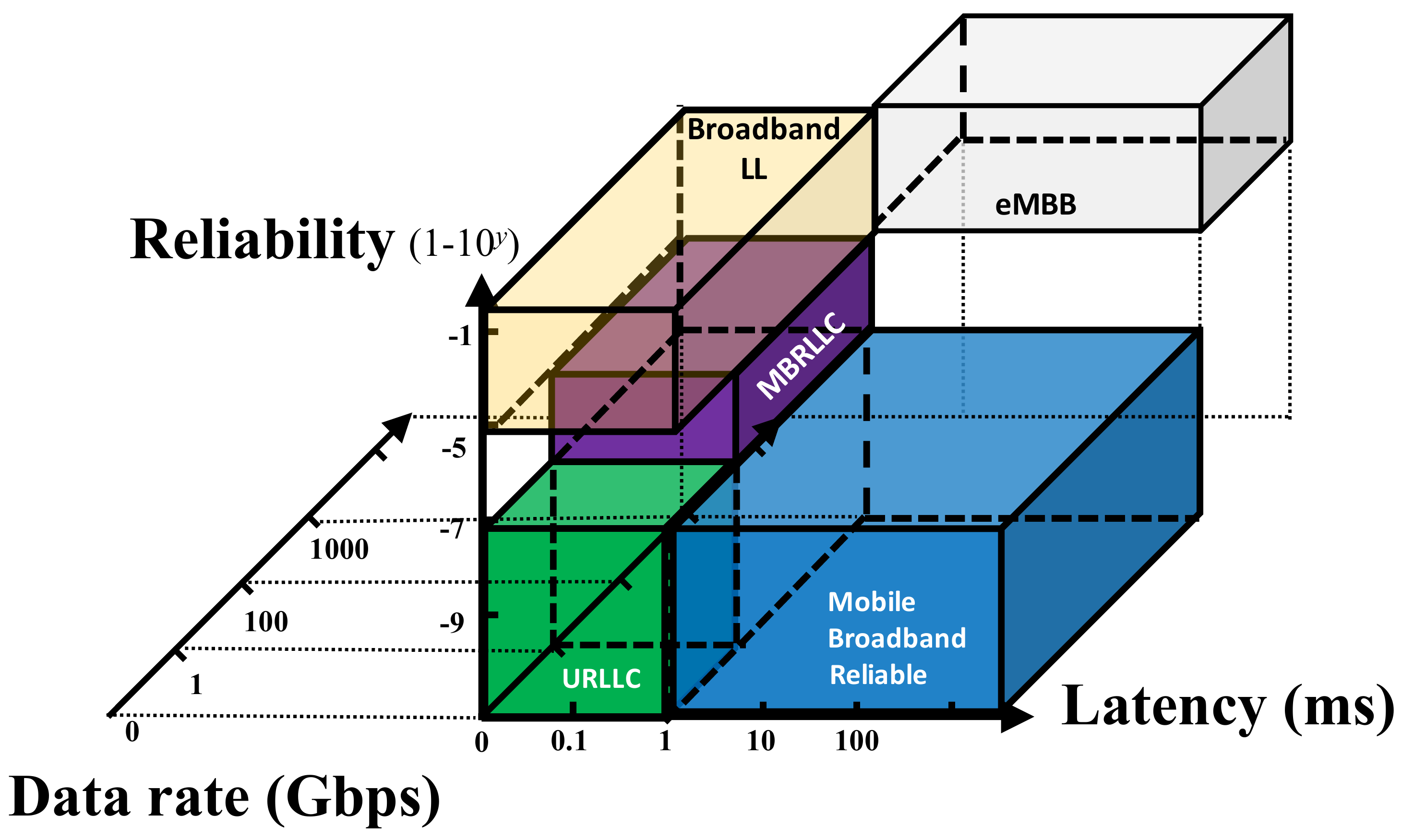}
    \vspace{-0.2cm}
    \caption{\label{figure2}MBRLLC services and several special cases (including classical eMBB and URLLC) within the rate-reliability-latency space. Other involved, associated metrics that are not shown include energy and network scale.}
  \end{center}\vspace{-0.7cm}
\end{figure*}

{\color{black}As evident from Section \ref{se:22}, the distinction between eMBB and URLLC will no longer be sustainable to support applications such as XR, wireless BCI, or CRAS. This is because these applications require, not only high reliability and low latency but also high 5G-eMBB-level data rates.} Hence, we propose a new service class called \emph{mobile broadband reliable low latency communication (MBRLLC)} that allows 6G systems to deliver any required performance within the rate-reliability-latency space. As seen in Fig. \ref{figure2}, MBRLLC generalizes classical URLLC and eMBB services. Energy efficiency is central for MBRLLC, not only because of its impact on reliability and rate, but also because of the resource-limited nature of 6G devices.  

\subsubsection{\bf{Massive URLLC}} 5G URLLC meant meeting reliability and latency of very specific uplink IoE applications such as smart factories, for which prior work \cite{Popovski-100} provided the needed fundamentals. However, 6G must scale classical URLLC across the device dimension thereby leading to a new \emph{massive URLLC (mURLLC)} service that merges 5G URLLC with legacy mMTC. mURLLC brings forth a reliability-latency-scalability  tradeoff which mandates a major departure from average-based network designs (e.g., average throughput/delay). Instead, a principled and scalable framework which accounts for delay, reliability, packet size, architecture, topology (across access, edge, and core) and decision-making under uncertainty is necessary \cite{MehdiMag}. 

\subsubsection{\bf{Human-Centric Services}} We propose a new class of 6G services, dubbed \emph{human-centric services (HCS)}, that require QoPE targets (tightly coupled with their human users, as explained in Section \ref{se:21}) rather than raw rate-reliability-latency metrics. Wireless BCI are a prime example of HCS in which network performance is determined by the physiology of the human users and their actions. For such services, a whole new set of QoPE metrics must be defined and offered as function of raw QoS and QoE metrics.

\subsubsection{\bf{Multi-Purpose 3CLS and Energy Services}} 6G systems must jointly deliver 3CLS services and their derivatives. They can also potentially offer energy to small devices via wireless energy transfer. Such \emph{multi-purpose 3CLS and energy services (MPS)} will be particularly important for applications such as CRAS. MPS require joint uplink-downlink designs and must meet target performance for the control (e.g., stability), computing (e.g., computing latency), energy (e.g., target energy to transfer), localization (e.g., localization precision),  as well as sensing and mapping functions (e.g., accuracy of a mapped radio environment).

 \begin{table*}
 \color{black}
\centering
  \newcommand{\tabincell}[2]{\begin{tabular}{@{}#1@{}}#2.4\end{tabular}}
\renewcommand\arraystretch{1.5}
 \caption{
    \vspace*{-0.05em}Summary of 6G service classes, their performance indicators, and example applications. }\label{ta:2}\vspace*{-0.5em}
\centering
\begin{tabular}{|c|l|l|}
\hline
 \multicolumn{1}{|c|}{\multirow{2}{*}{\textbf{Service}}}&\multicolumn{1}{|c|}{\multirow{2}{*}{\textbf{Performance Indicators
}}}   &  \multicolumn{1}{|c|}{\multirow{2}{*}{\textbf{ Example Applications}}}  \\ 
 &&\\
\hline
\multirow{4}{*}{{\bf MBRLLC}}& \multirow{1}{*}{{$\bullet$ Stringent rate-reliability-latency  requirements.}} &\multirow{1}{*}{$\bullet$ \color{black}XR/AR/VR.}\\
&$\bullet$ Energy efficiency.&$\bullet$ \color{black}{Autonomous vehicular systems}. \\
&$\bullet$ Rate-reliability-latency in mobile environments.&$\bullet$ \color{black}{Autonomous drones}. \\ 
&&$\bullet$ \color{black}{Legacy eMBB and URLLC}.\\
\hline

\multirow{4}{*}{{\bf mURLLC}}& \multirow{1}{*}{{$\bullet$ Ultra high reliability.}} &\multirow{1}{*}{$\bullet$ Classical Internet of Things.}\\
&$\bullet$ Massive connectivity.&$\bullet$ User tracking.\\
&$\bullet$ Massive reliability.&$\bullet$ Blockchain and DLT. \\ 
&$\bullet$ Scalable URLLC.
&\multirow{1}{*}{$\bullet$ Massive sensing.} \\ 
&&\multirow{1}{*}{$\bullet$ Autonomous robotics.}  \\ 
\hline

\multirow{4}{*}{{\bf HCS}}& \multirow{2}{5cm}{{$\bullet$ QoPE capturing raw wireless metrics as well as human and physical factors.}} &\multirow{1}{*}{$\bullet$ BCI.} \\
&&$\bullet$ Haptics.\\
&&$\bullet$ Empathic communication. \\ 
&& $\bullet$  Affective communication.  \\ 
\hline
\multirow{6}{*}{{\bf MPS}}& \multirow{1}{*}{{$\bullet$ Control stability.}} &\multirow{1}{4cm}{$\bullet$ CRAS.}\\
& $\bullet$ Computing latency. &$\bullet$ Telemedicine.\\
& $\bullet$ Localization accuracy.&$\bullet$ Environmental mapping and imaging.\\ 
&$\bullet$ Sensing and mapping accuracy. &$\bullet$ Some special cases of XR services. \\ 
&$\bullet$ Latency and reliability for communications. & \\ 
&$\bullet$ Energy. & \\ 

\hline
\end{tabular}
 \vspace{-0.2cm}

\end{table*}

\footnotetext[3]{\color{black}{Here, spectral and energy efficiency gains are captured by the concept of area spectral and energy efficiency.}}

\section{6G: Enabling Technologies}\label{se:3}
To enable the aforementioned services and guarantee their performance, a cohort of new, disruptive technologies must be integrated into 6G.

\subsubsection{\bf{Above 6 GHz for 6G -- from Small Cells to Tiny Cells}} As per Trends 1 and 2, the need for higher data rates and SEE anywhere, anytime in 6G motivates exploring higher frequency bands beyond sub-6 GHz. As a first step, this includes further developing mmWave technologies to make \emph{mobile mmWave} a reality in early 6G systems. As 6G progresses, exploiting frequencies beyond mmWave, at the terahertz (THz) band, will become necessary \cite{Rappaport-6G}. To exploit higher mmWave and THz frequencies, the size of the 6G cells must shrink from small cells to ``tiny cells'' whose radius is only few tens meters. This motivates new architectural designs that need much denser deployments of  tiny cells and new high-frequency mobility management techniques.

\subsubsection{\bf{Transceivers with Integrated Frequency Bands}} On their own, dense high-frequency tiny cells may not be able to provide the seamless connectivity required for mobile 6G services. Instead, an integrated system that can leverage multiple frequencies across the microwave/mmWave/THz spectra (e.g., using multi-mode base stations) is needed to provide seamless connectivity at both wide and local area levels.

\subsubsection{\bf{Communication with Large Intelligent Surfaces}} Massive MIMO will be integral to both 5G and 6G due to the need for better SEE, higher data rates, and higher frequencies (Trend 1). However, for 6G systems, as per Trend 3, we envision an initial leap from traditional massive MIMO towards large intelligent surfaces (LISs) and smart environments \cite{LIS-1}  that can provide massive surfaces for wireless communications and for heterogeneous devices (Trend 7). LISs enable innovative ways for communication such as by using holographic radio frequency (RF) and holographic MIMO.

\subsubsection{\bf{Edge AI}}  AI is witnessing an unprecedented interest from the wireless community \cite{Chen-AI} driven by recent breakthroughs in deep learning, the increase in available data (Trend 4), and the rise of smart devices (Trend 7). Imminent 6G use cases for AI (particularly for reinforcement learning) revolve around creating SSNs (Trend 5) that can autonomously sustain high KPIs and manage resources, functions, and network control. AI will also enable 6G to automatically provide MPS to its users and to send and create 3D radio environment maps (Trend 6). These short-term AI-enabled 6G functions will be complemented by a so-called ``collective network intelligence'' in which  network intelligence is pushed at the edge, running AI and learning algorithms on edge devices (Trend 7) to provide distributed autonomy. This new edge AI leap will create a 6G system that can integrate the services of Section \ref{se:2}, realize 3CLS, and potentially replace classical frame structures. 

\subsubsection{\bf{Integrated Terrestrial, Airborne, and Satellite Networks}} Beyond their inevitable role as 6G users, drones can be leveraged to complement terrestrial networks by providing connectivity to hotspots and to areas with scarce infrastructure. Meanwhile, both drones and terrestrial base stations may require satellite connectivity with low orbit satellites (LEO) and CubeSats to provide backhaul support and additional wide area coverage. Integrating terrestrial, airborne, and satellite networks \cite{Mohammad-2019} and \cite{Halim} into a single wireless system will be essential for 6G.

\subsubsection{\bf{Energy Transfer and Harvesting}} 6G could be the cellular system that can provide energy, along with 3CLS (Trend 6). As wireless energy transfer is maturing, we foresee 6G base stations providing basic power transfer for devices, particularly implants and sensors (Trend 7). Adjunct energy-centric ideas, such as energy harvesting and backscatter will also be a component of 6G.

\subsubsection{\bf{Beyond 6G}} A handful of technologies will mature along the same time of 6G and, hence, potentially play a role towards the end of the 6G standardization and research process. One prominent example is \emph{quantum computing and communications} that can provide security and long-distance networking. Currently, major research efforts are focused on the quantum realm and we expect them to intersect with 6G. Other similar beyond 6G technologies include integration of RF and non-RF links (including optical, neural, molecular, and other channels).

\section{6G: Research Agenda and Open Problems}\label{se:4}

\begin{figure*}[!t]
  \begin{center}
   \vspace{0cm}
    \includegraphics[width=14cm]{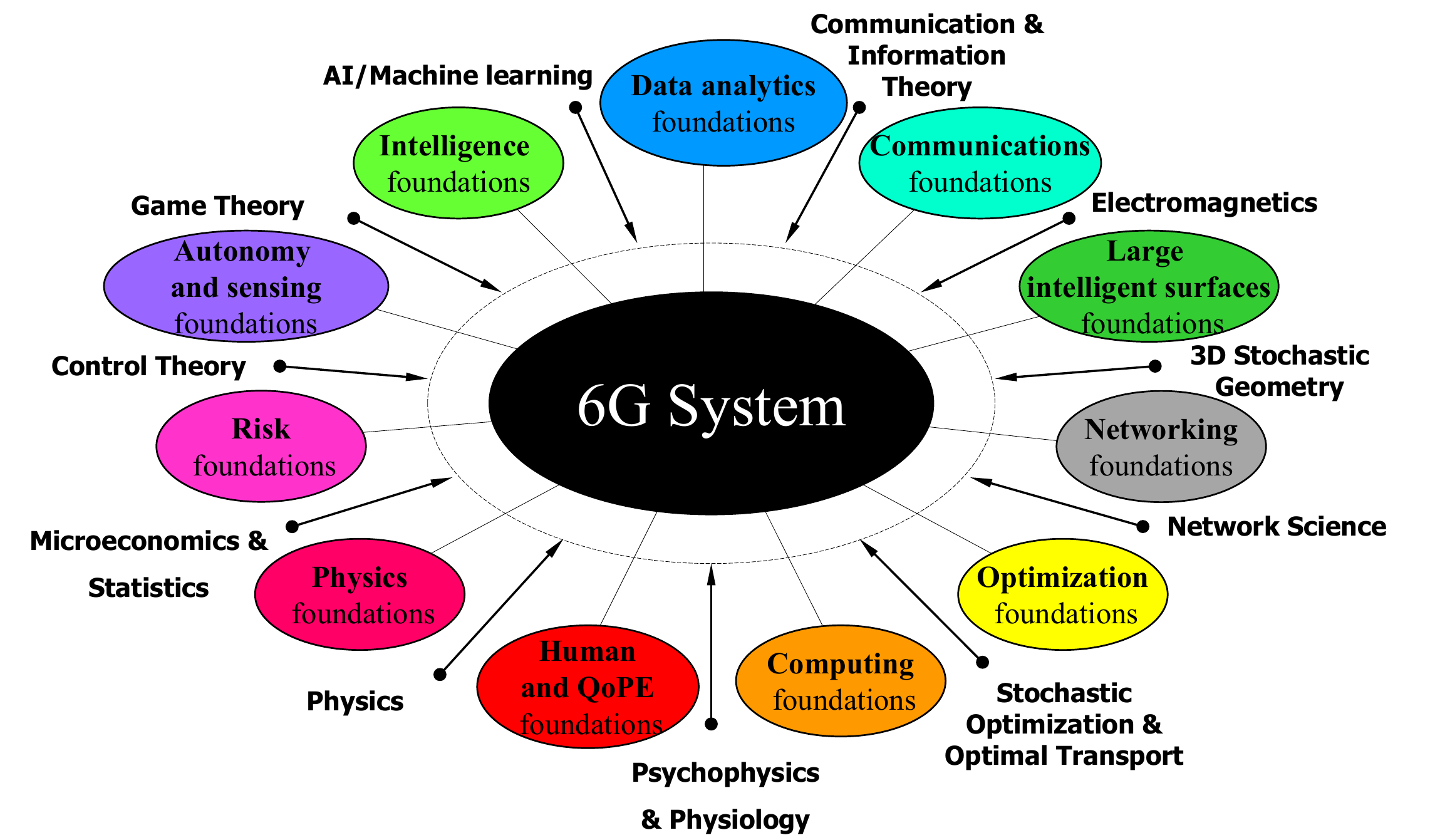}
    \vspace{-0.2cm}
    \caption{\label{figure3} Necessary foundations and associated analytical tools for 6G.}
  \end{center}\vspace{-0.7cm}
\end{figure*}

Building on the identified trends in Section \ref{se:2} and the enabling technologies in Section \ref{se:3}, we now put forward a research agenda for 6G (summarized in Table \ref{ta:3}). 

\subsubsection{\bf{3D Rate-Reliability-Latency Fundamentals}} Fundamental 3D performance of 6G systems, in terms of rate-reliability-latency tradeoffs and SEE is needed. Such analysis must quantify the spectrum, energy, and communication requirements that 6G needs in order to support the identified driving applications. Recent works in \cite{MehdiMag} and \cite{Ali-ICC} provide a first step in this direction.

\subsubsection{\bf{Exploring Integrated, Heterogeneous High-Frequency Bands}} Exploiting mmWave and THz in 6G brings forth several new open problems. For mmWave, supporting high mobility at mmWave frequencies will be a central open problem. For THz, new transceiver architectures and propagation models are needed \cite{Rappaport-6G}. High power, high sensitivity, and low noise figure are key transceiver features needed to overcome the very high THz path loss. {\color{black}Once these physical layer aspects are well-understood, there is a need to develop new network and link-layer protocols to optimize the use of cross-frequency resources while taking into account the highly varying and uncertain nature of the mmWave and THz environments.} Another important direction is to study the co-existence of THz, mmWave, and microwave cells across all layers, building on early works such as \cite{Omid-2019}. 

\subsubsection{\bf{3D  Networking}} Due to the integration of ground and airborne networks, as outlined in Section \ref{se:3}, 6G must support communications in 3D space, including serving users in 3D and deploying 3D base stations (e.g., tethered balloons or temporary drones). This requires concerted research on various fronts. First, measurement and (data-driven) modeling of the 3D propagation environment is needed. Second, new approaches for 3D frequency and network planning (e.g., where to deploy base stations, tethered balloons, or even drone-base stations) must be developed. Our work in \cite{Mohammad-2019} showed that such 3D planning is substantially different from conventional 2D networks due to the new altitude dimension and the associated degrees of freedom. Finally, new network optimizations for mobility management, routing, and resource management in 3D are needed.

\subsubsection{\bf{Communication with LIS}} As per Trend 3, 6G will provide wireless connectivity via smart LIS environments that include active frequency selective surfaces, metallic passive reflectors, passive/active reflect arrays, as well as nonreconfigurable and reconfigurable metasurfaces. Open problems here range from the optimized deployment of passive reflectors and metasurfaces to AI-powered operation of LISs. Fundamental analysis to understand the performance of smart surfaces, in terms of rate, latency, reliability, and coverage is needed, building on the early work in \cite{LIS-1}. Another important direction is to investigate the potential of using LIS-based reflective surfaces to enhance the range and coverage of tiny cells and to dynamically modify the propagation environment.  

\subsubsection{\bf{AI for Wireless}}  AI brings forward many major research directions for 6G. Beyond the need for massive, small data analytics as well as using machine learning (ML) and AI-based SSNs (realized using reinforcement learning and game theory), there is also a need to operate ML algorithms reliably over 6G to deliver the applications of Section \ref{se:2}. To perform these critical application tasks, low-latency, high-reliability and scalable AI is needed, along with a reliable infrastructure \cite{Chen-AI} and \cite{MehdiEdge}. This joint design of ML and wireless networks is a key 6G research area. 

\subsubsection{\bf{QoPE Metrics}} The design of QoPE metrics that integrate physical factors from human physiology (for HCS services) or from a control system (for CRAS) is an important 6G research area, especially in light of new, emerging devices (Trend 7). This requires both real-world psychophysics experiments as well as new, rigorous mathematical expressions for QoPE that combine QoS, QoE, and human perceptions. Theoretical development of QoPE can be achieved using techniques from other disciplines such as operations research (e.g., multi-attribute utility theory (see \cite{Chen-Saad-VR-2018})) and machine learning (see \cite{kasgari2018human}). 6G will be the first generation to enable a new breed of applications (wireless BCI) leveraging multiple human cognitive senses.

\subsubsection{\bf{ Joint Communication and Control}} 6G needs to pervasively support CRAS. The performance of  CRAS is governed by real-world control systems whose operation requires data input from wireless 6G links. Therefore, operating CRAS over 6G systems requires a \emph{communication and control co-design}, whereby the performance of the 6G wireless links is optimized to cater for the stability of the control system and vice versa. Due to the traditional radio-centric focus (3GPP and IEEE fora), such a co-design has been overlooked in 5G. Meanwhile, prior works on networked control abstract the specifics of the wireless network and cannot apply to cellular communications. This makes the communication-control co-design a key research topic in 6G.

\subsubsection{\bf{3CLS}} The idea of joint communication and control must be extended to the joint design of the entire 3CLS functions. The interdependence between computing, communication, control, localization, sensing, energy, and mapping has not yet been fully explored in an end-to-end manner. Key questions range from how to jointly meet the performance of all 3CLS services to multi-modal sensor fusion for reconstructing 3D images and navigating in unknown environments for navigating robots, autonomous driving, etc. 3CLS is needed for various applications including CRAS, XR, and DLT.

\subsubsection{\bf{\color{black}6G Protocol Designs}}{\color{black}Owing to all trends discussed in Section \ref{se:22} and their challenges, compared to 5G, 6G will require radical new protocol designs. For instance, 6G must introduce new, AI-driven protocols for signaling, scheduling, and coordination that can replace conventional 5G protocols that rely on pre-determined network parameters and rigid frame structures. These new 6G protocols will, in contrast, continuously adapt to the current and projected state of the wireless environment. As 6G evolves, there will be a need for new, dynamic multiple access \cite{8663993} protocols that can dynamically change the type of multiple access (orthogonal or non-orthogonal, random or scheduled) used depending on the needs of the applications and the network state. Moreover, novel handover protocols must be designed to account for the 3D nature of the 6G system and the presence of different types of mobile devices. New protocols for authentication and identification will also be needed to handle the new breed of wireless devices that include  drones, vehicles, as well as embedded and implanted devices. Finally, all 6G protocols must be distributed and able to leverage datasets distributed across the network edge.}

\subsubsection{\bf{RF and non-RF Link Integration}} 6G will witness a convergence of RF and non-RF links that encompass optical, visible light communication (VLC), molecular communication, and neuro-communication, among others. Design of such joint RF/non-RF systems is an open research area.

\subsubsection{\bf{Holographic Radio}} RF holography (including holographic MIMO) and spatial spectral holography can be made possible with 6G due to the use of LIS and similar structures. Holographic RF allows for control of the entire physical space and the full closed loop of the electromagnetic field through spatial spectral holography and spatial wave field synthesis. This greatly improves spectrum efficiency and network capacity, and helps the integration of imaging and wireless communication. How to realize holographic radio is a widely open area. 

An overview on the necessary analytical tools and fundamentals related to these open problems is shown in Fig. \ref{figure3}.

 \begin{table*}
\centering
  \newcommand{\tabincell}[2]{\begin{tabular}{@{}#1@{}}#2.4\end{tabular}}
\renewcommand\arraystretch{1.2}
 \caption{
    \vspace*{-0.05em}Summary of Research Areas }\label{ta:3}\vspace*{-0.5em}
\centering
\begin{tabular}{|c|l|l|}
\hline
 \multicolumn{1}{|c|}{\multirow{2}{*}{\textbf{Research Area}}}&\multicolumn{1}{|c|}{\multirow{2}{*}{\textbf{Challenges
}}}   &  \multicolumn{1}{|c|}{\multirow{2}{*}{\textbf{ Open Problems}}}  \\ 
 &&\\
\hline
\multirow{4}{3.5cm}{{\bf3D Rate-Reliability-Latency Fundamentals}}& \multirow{1}{*}{{$\bullet$ Fundamental communication limits.}} &\multirow{1}{*}{$\bullet$ 3D performance analysis of rate-reliability-latency region.}\\
&$\bullet$ 3D nature of 6G systems.&$\bullet$ Characterization of achievable  rate-reliability-latency targets. \\
&&$\bullet$ 3D SEE characterization. \\ 
&&   \multirow{2}{6cm}{{$\bullet$ Characterization of energy and spectrum needs for rate-reliability-latency targets.}}  \\
&&\\
\hline

\multirow{7}{3.5cm}{{\bf Exploring Integrated, Heterogeneous High-Frequency Bands}}& \multirow{1}{*}{{$\bullet$ Challenges of operation in highly mobile systems.  }} &\multirow{1}{*}{$\bullet$ Effective mobility management for mmWave and THz systems.}\\
&$\bullet$ Susceptibility to blockage.&$\bullet$ Cross-band physical, link, and network layer optimization. \\
&$\bullet$ Short range.  &$\bullet$ Coverage and range improvement. \\ 
&$\bullet$ Lack of propagation models. &\multirow{1}{*}{$\bullet$ Design of mmWave and THz tiny cells.} \\ 
&$\bullet$ Need for high fidelity hardware.   &\multirow{1}{*}{$\bullet$ Design of new high fidelity hardware for THz.}  \\ 
&$\bullet$ Co-existence of frequency bands.&\multirow{2}{7cm}{$\bullet$  Propagation measurements and modeling across mmWave and THz bands.}  \\ 
&&\\
\hline
\multirow{3}{*}{{\bf 3D Networking}}& \multirow{1}{*}{{$\bullet$ Presence of users and base stations in 3D.}} &\multirow{1}{*}{$\bullet$ 3D propagation modeling.}\\
&$\bullet$ High mobility.&$\bullet$ 3D performance metrics.\\
&&$\bullet$ 3D mobility management and network optimization. \\ 
\hline

\multirow{7}{*}{{\bf Communication with LIS}}& \multirow{1}{*}{{$\bullet$ Complex nature of LIS surfaces.}} &\multirow{1}{*}{$\bullet$ Optimal deployment and location of LIS surfaces.}\\
&$\bullet$ Lack of existing performance models.&$\bullet$ LIS reflectors vs. LIS base stations.\\
&$\bullet$ Lack of propagation models.&$\bullet$ LIS for energy transfer. \\ 
&$\bullet$ Heterogeneity of 6G devices and services.&\multirow{1}{*}{$\bullet$ AI-enabled LIS.} \\ 
&  \multirow{2}{5cm}{{$\bullet$ Ability of LIS to provide different functions (reflectors, base stations, etc.).}}   &\multirow{1}{*}{$\bullet$ LIS across 6G services.}  \\ 
&&  \multirow{2}{7cm}{{$\bullet$ Fundamental performance analysis of LIS transmitters and reflectors at various frequencies.}} \\
&&\\
\hline

\multirow{4}{*}{{\bf AI for Wireless}}& \multirow{1}{*}{{$\bullet$ Design of low-complexity AI solutions.}} &\multirow{1}{*}{$\bullet$ Reinforcement learning for SON.}\\
&$\bullet$ Massive, small data.&$\bullet$ Big and small data analytics.\\
&&$\bullet$ AI-powered network management. \\ 
& &\multirow{1}{*}{$\bullet$ Edge AI over wireless systems.}  \\ 
\hline

\multirow{4}{*}{{\bf New QoPE Metrics}}& \multirow{1}{*}{{$\bullet$ Incorporate raw metrics with human perceptions.}} &\multirow{1}{*}{$\bullet$ Theoretical development of QoPE metrics.}\\
&  \multirow{1}{5cm}{{$\bullet$ Accurate modeling of human perceptions and physiology.}}&$\bullet$ Empirical QoPE characterization.\\
&&$\bullet$ Real psychophysics experiments.\\ 
& &$\bullet$ Definition of realistic QoPE targets and measures. \\ 
\hline

\multirow{4}{*}{{\bf Joint Communication and Control}}& \multirow{1}{*}{{$\bullet$ Integration of control and communication metrics.}} &\multirow{1}{*}{$\bullet$ Communication and control systems co-design.}\\
&  \multirow{1}{*}{{$\bullet$ Handling dynamics and multiple time scales.}}&$\bullet$ Control-enabled wireless metrics.\\
&&$\bullet$ Wireless-enabled control metrics.\\ 
& &$\bullet$ Joint optimization for CRAS. \\ 
\hline

\multirow{4}{*}{{\bf 3CLS }}& \multirow{1}{*}{{$\bullet$ Integration of multiple functions.}} &\multirow{1}{*}{$\bullet$ Design of 3CLS metrics.}\\
&  \multirow{1}{*}{{$\bullet$ Lack of prior models.}}&$\bullet$ Joint 3CLS optimization.\\
&&$\bullet$ AI-enabled 3CLS.\\ 
& &$\bullet$ Energy efficient 3CLS. \\ 
\hline

\multirow{10}{*}{{\bf \color{black} 6G Protocol Designs }}& \multirow{2}{5.2cm}{{\color{black}$\bullet$ 6G protocols must operate in 3D space and across different propagation environments.}} &\multirow{3}{7cm}{{\color{black}$\bullet$ Design of scheduling, coordination, and signaling protocols that do not require pre-determined, rigid frame structures.}}\\
& &\\
&\multirow{2}{5.2cm}{{\color{black}$\bullet$ Presence of heterogeneous devices with different capabilities and mobility patterns.}}&\\ 
& &\multirow{1}{7cm}{{\color{black}$\bullet$	Development of adaptive multiple access protocols.}} \\ 
&\multirow{2}{5.2cm}{{\color{black}$\bullet$ Need for protocols that can learn and adapt to the environment.}}&\multirow{2}{7cm}{{\color{black}$\bullet$	Design of proactive and dynamic handover mechanisms that can cope with different mobility patterns in 3D space.}} \\ 
& &\\
&&\multirow{2}{7cm}{{\color{black}$\bullet$ Introduction of new authentication and identification protocols tailored to 6G devices.}} \\ 
& &\\
&&\multirow{2}{7cm}{{\color{black}$\bullet$ Design of distributed, edge AI-inspired protocols for executing multiple 6G functions.} } \\ 
& &\\
\hline

\multirow{3}{*}{\bf  RF and non-RF Link Integration}& \multirow{1}{*}{{$\bullet$ Different physical nature of RF/non-RF interfaces.}} &\multirow{1}{*}{$\bullet$ Design of joint RF/non-RF hardware.}\\
& &$\bullet$ System-level analysis of joint RF/non-RF systems.\\
&&$\bullet$ Use of RF/non-RF systems for various 6G services.\\ 
\hline

\multirow{4}{*}{{\bf Holographic Radio }}& \multirow{1}{*}{{$\bullet$ Lack of existing models.}} &\multirow{1}{*}{$\bullet$ Design of holographic MIMO using LIS.}\\
&  \multirow{1}{*}{{$\bullet$ Hardware and physical layer challenges.}}&$\bullet$ Performance analysis of holographic RF.\\
&&$\bullet$ 3CLS over holographic radio.\\ 
& &$\bullet$ Network optimization with holographic radio. \\ 
\hline

\end{tabular}
 \vspace{-0.2cm}

\end{table*}

\section{Conclusion and Recommendations}\label{se:5}
This article laid out a bold new vision for 6G systems that outlines the trends, challenges, and associated research. While many topics will come as a natural 5G evolution, new avenues of research such as LIS communication, 3CLS, holographic radio, and others will create an exciting research agenda for the next decade. We conclude with several recommendations:

\begin{itemize}
\item {\bf \underline {Recommendation 1}:} A first step towards 6G is to enable MBRLLC and mobility management at high-frequency mmWave bands and beyond (i.e., THz).
\item {\bf \underline {Recommendation 2}:} 6G requires a move from radio-centric system design (\`a-la-3GPP) towards an end-to-end 3CLS co-design under the orchestration of an AI-driven intelligence substrate.
 \item {\bf \underline {Recommendation 3}:}  The 6G vision will not be a simple case of exploring additional, high-frequency spectrum bands to provide more capacity. Instead, it will be driven by a diverse portfolio of  applications, technologies, and techniques (see Figs. \ref{figure1} and \ref{figure3}). 
 \item {\bf \underline {Recommendation 4}:}   6G will transition from the smartphone-base station paradigm into a new era of smart surfaces communicating with human-embedded implants.
 \item {\bf \underline {Recommendation 5}:}   Performance analysis and optimization of 6G requires operating in 3D space and moving away from simple averaging towards fine-grained analysis that deals with tails, distributions, and QoPE.
\end{itemize}

\bibliographystyle{IEEEbib}
\def\baselinestretch{0.91}
\bibliography{references}

\end{document}